%% file: hcha_189.tex
\documentclass[12pt,a4paper,dvips]{article}
\usepackage{a4p}
\usepackage{cite,mcite}
\usepackage{epsfig}   
\usepackage{graphicx}
\usepackage{physics }
\usepackage{l3_title,ifthen,Lep}
\journalname{Phys. Lett. B}

\date{August 26, 1999}
\preprint{99-120}
\newlength{\capindent}
\newlength{\capwidth}
\newlength{\figwidth}
\newcommand{\icaption}[2][!*!,!]{\hspace*{\capindent}%
  \begin{minipage}{\capwidth}
    \ifthenelse{\equal{#1}{!*!,!}}%
      {\caption{#2}}%
      {\caption[#1]{#2}}
  \end{minipage}}
\def\tntn{\mathrm{\tau^+\nu_\tau\tau^-\nbar_\tau}}
\def\cstn{\mathrm{c\bar{s} \tau^-\nbar_\tau}}
\def\cscs{\mathrm{c\bar{s} \bar{c}s}}

\def\HHtntn{\mathrm{\Hp\Hm \ra \tntn}}
\def\HHcstn{\mathrm{\Hp\Hm \ra \cstn}}
\def\HHcscs{\mathrm{\Hp\Hm \ra \cscs}}
\def\l{\ifmath{\mathrm{\ell}}}

\begin{document}

\begin{titlepage}

\mathversion{bold}
\title{Search for Charged Higgs Bosons in $\bf\ee$ Collisions
       at $\sqrt{s} = 189$ \GeV{}}

\author{The L3 Collaboration}

\mathversion{normal}

\begin{abstract}
  
  A search for pair--produced charged Higgs bosons is performed with the
  L3 detector at LEP using data collected at a  centre--of--mass  energy
  of 188.6~\GeV{},  corresponding  to an integrated  luminosity of 176.4
  \pb.  Higgs  decays  into a charm  and a  strange  quark or into a tau
  lepton  and its  associated  neutrino  are  considered.  The  observed
  events  are  consistent  with the  expectations  from  Standard  Model
  background  processes.  A lower limit of  65.5~\GeV{}  on the  charged
  Higgs mass is derived at 95\%  confidence  level,  independent  of the
  decay branching ratio $\mathrm{Br(H^\pm\ra \tau\nu)}$.

\end{abstract}

\submitted

\end{titlepage}


\section*{Introduction}

In    the     Standard     Model~\cite{standard_model},     the    Higgs
mechanism~\cite{higgs_mech}  requires  one  doublet  of  complex  scalar
fields which leads to the  prediction of a single  neutral  scalar Higgs
boson.  Extensions to the minimal  Standard  Model contain more than one
Higgs   doublet~\cite{higgs_hunter}.  In  particular,  models  with  two
complex Higgs doublets  predict two charged Higgs bosons  ($\Hpm$).  The
discovery  of a charged  Higgs  particle  would be evidence  for physics
beyond the Standard Model.

The  search  is  performed  in  the  three  decay  channels   $\HHtntn$,
$\HHcstn$\footnote{The  charge conjugate reaction is included throughout
the  letter.}  and  $\HHcscs$.  This  allows the  interpretation  of the
results to be independent  of the branching  ratio  $\mathrm{Br(H^\pm\ra
\tau\nu)}$.  

The results  include and supersede the previous  lower limit on the mass
of       charged        Higgs        bosons        established        by
L3~\cite{chhiggs_130_180,l3_48_50}.  Results from other LEP  experiments
are published in Reference~\cite{other_lep}.


\section*{Data Analysis}

This letter describes the search for pair--produced charged Higgs bosons
using the data collected  with the L3  detector~\cite{l3_det}  at LEP at
$\sqrt{\mathrm{s}}=188.6$~\GeV,    corresponding    to   an   integrated
luminosity of 176.4  pb$^{-1}$.  The analyses  remain  almost  unchanged
since our previous publication at centre--of--mass  energies between 130
and   183~\GeV~\cite{chhiggs_130_180},   because  they  show  a  similar
performance.

The charged Higgs cross section is calculated using the HZHA Monte Carlo
program~\cite{hzha}.  For   the   efficiency   estimates,   samples   of
$\mathrm{\ee\ra  (Z/\gamma)  \ra\Hp\Hm}$  events are generated  with the
PYTHIA Monte Carlo  program~\cite{jetset73}  for Higgs masses between 50
and  90~\GeV{}  in mass steps of  5~\GeV{}.  About 1000  events for each
final  state  are  generated  at each  Higgs  mass.  For the  background
studies  the  following  Monte  Carlo  generators  are used:  PYTHIA for
$\ee\ra\qqbar(\gamma)$,      $\ee\ra\Zo\Zo$     and      $\ee\ra\Zo\ee$,
KORALW~\cite{KORALW}   for  $\ee\ra\Wp\Wm$,   PHOJET~\cite{PHOJET}   for
$\ee\ra\ee\qqbar$,   DIAG36~\cite{DIAG36}  for   $\ee\ra\ee\ell^+\ell^-$
($\l=\e,\mu,\tau$),  KORALZ~\cite{KORALZ40}  for  $\ee\ra\mu^+\mu^-$ and
$\ee\ra\tau^+\tau^-$ and BHAGENE3~\cite{BHAGENE3}  for $\ee\ra\ee$.  The
L3    detector     response    is    simulated     using    the    GEANT
program~\cite{my_geant}  which takes into  account the effects of energy
loss, multiple scattering and showering in the detector.

\subsection*{{\boldmath{Search in the $ \HHtntn$ Channel}}}

\begin{figure}[hp]
\epsfig{figure=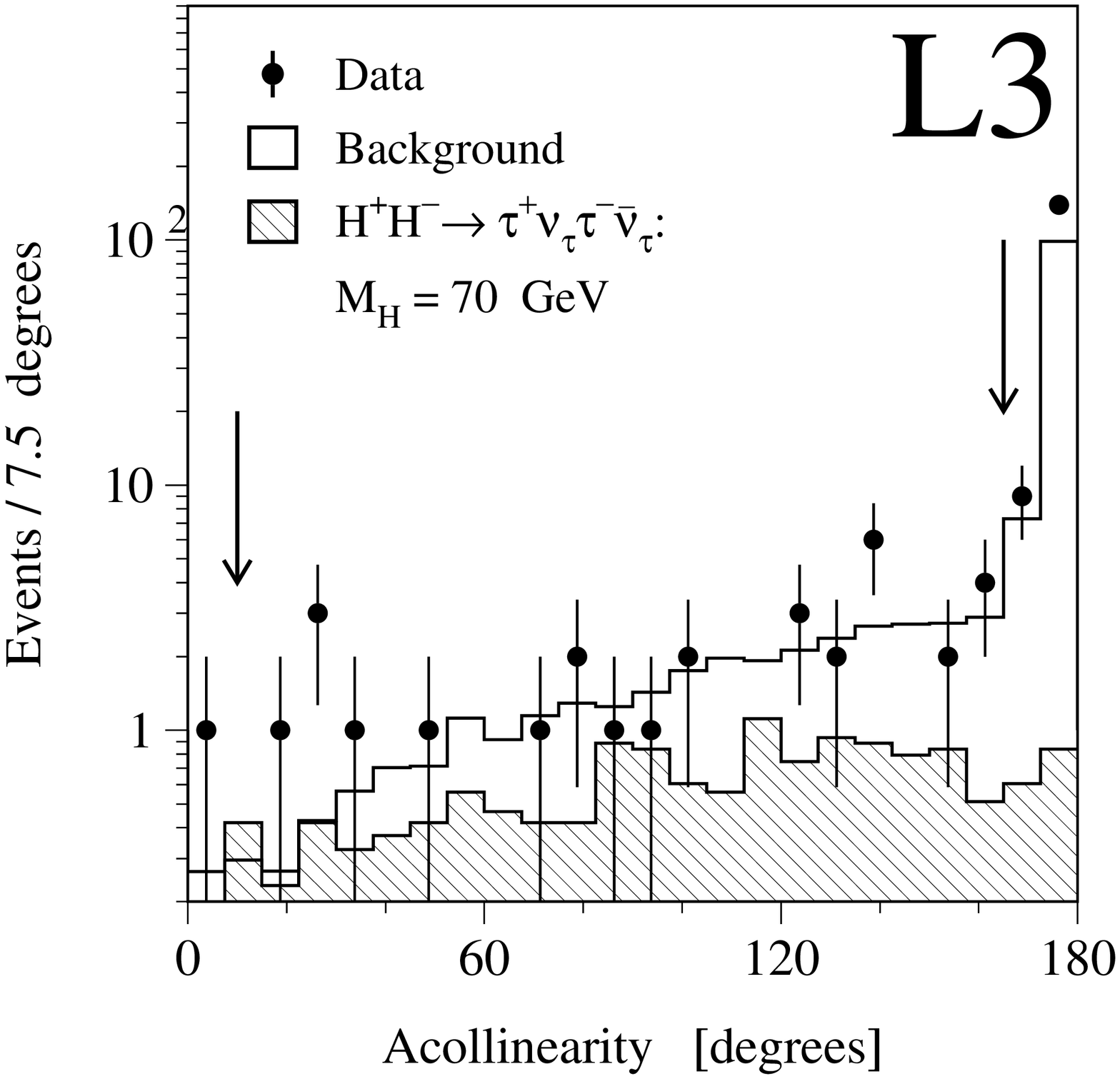,width=0.95\textwidth}
\caption[]{\label{fig:lepton}  The  maximum  angle  between  any pair of
  tracks, or  acollinearity,  for the $\HHtntn$  channel after all other
  cuts have been applied.  The arrows show the position of the cut.  The
  hatched histogram indicates the expected  distribution for a 70~\GeV{}
  Higgs at $\mathrm{Br(H^\pm\ra\tau\nu)} = 1$.}
\end{figure}

The  signature  for the leptonic  decay channel is a pair of tau leptons
with large missing energy and momentum, giving rise to low  multiplicity
events with low visible  energy and uniform  acollinearity,  the maximum
angle  between  any  pair of  tracks.  This  distribution  is  shown  in
Figure~\ref{fig:lepton}.  The   performance   of  the  analysis  is  not
affected  by  the  increased  centre--of--mass  energy,  and  the  event
selection       remains       unchanged       since       our       last
publication~\cite{chhiggs_130_180}.

The  efficiency of the  $\HHtntn$  selection for several Higgs masses is
shown in  Table~\ref{eff}.  The total number of events  selected in data
is 30, where 32.5  background  events are expected from  Standard  Model
processes.  Almost all of the  remaining  background  comes  from W pair
production.

\subsection*{{\boldmath{Search in the $\HHcstn$ Channel}}}

The semileptonic  final state $\HHcstn$ is characterised by two hadronic
jets, a tau lepton and missing  momentum.  The  selection  criteria  are
slightly modified with respect to the analysis at lower centre--of--mass
energies~\cite{chhiggs_130_180},  in order to gain sensitivity at higher
masses.

\begin{figure}[hp]
$$ \mbox{\epsfig{figure=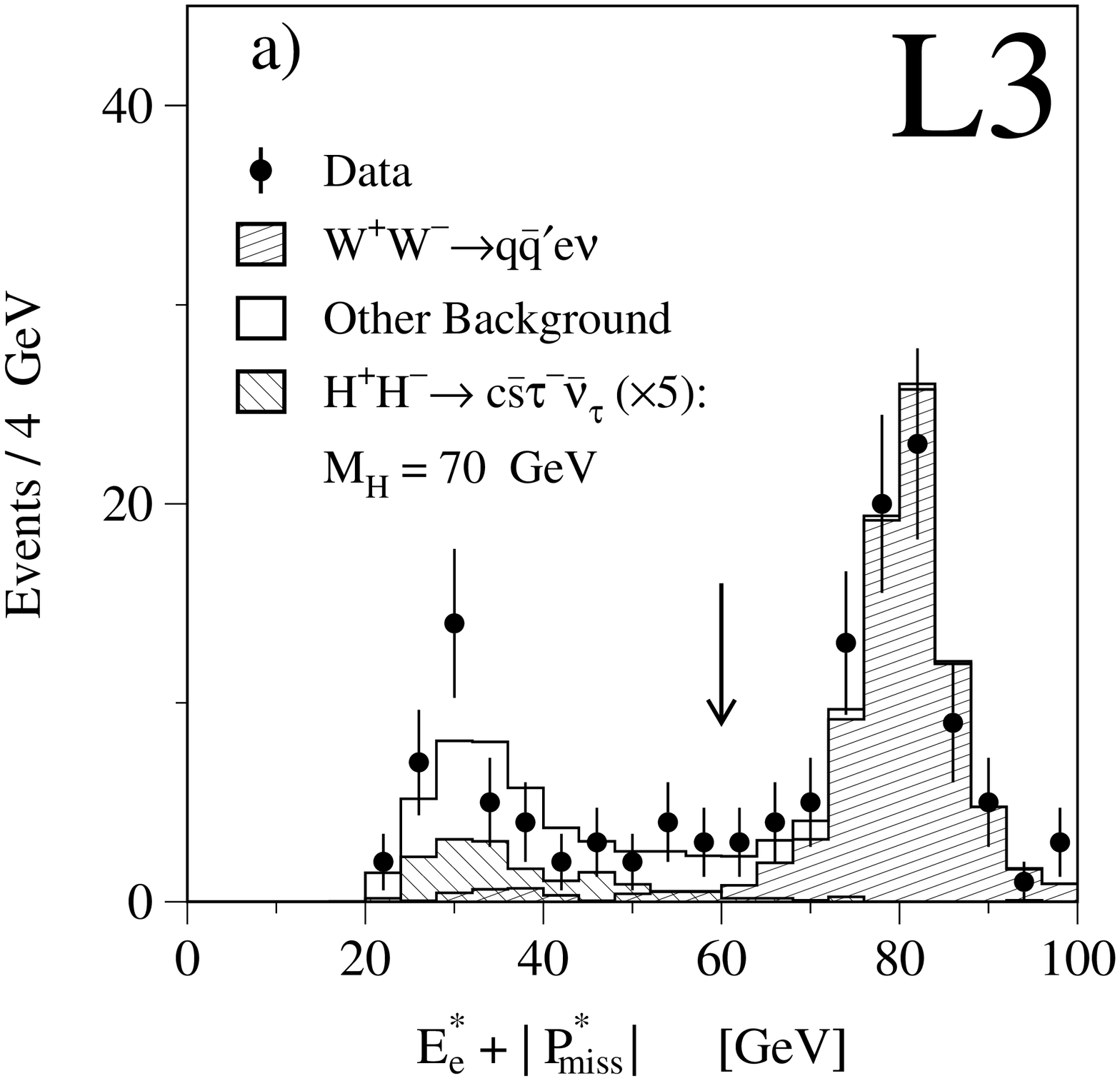,width=0.65\textwidth}} $$
$$ \mbox{\epsfig{figure=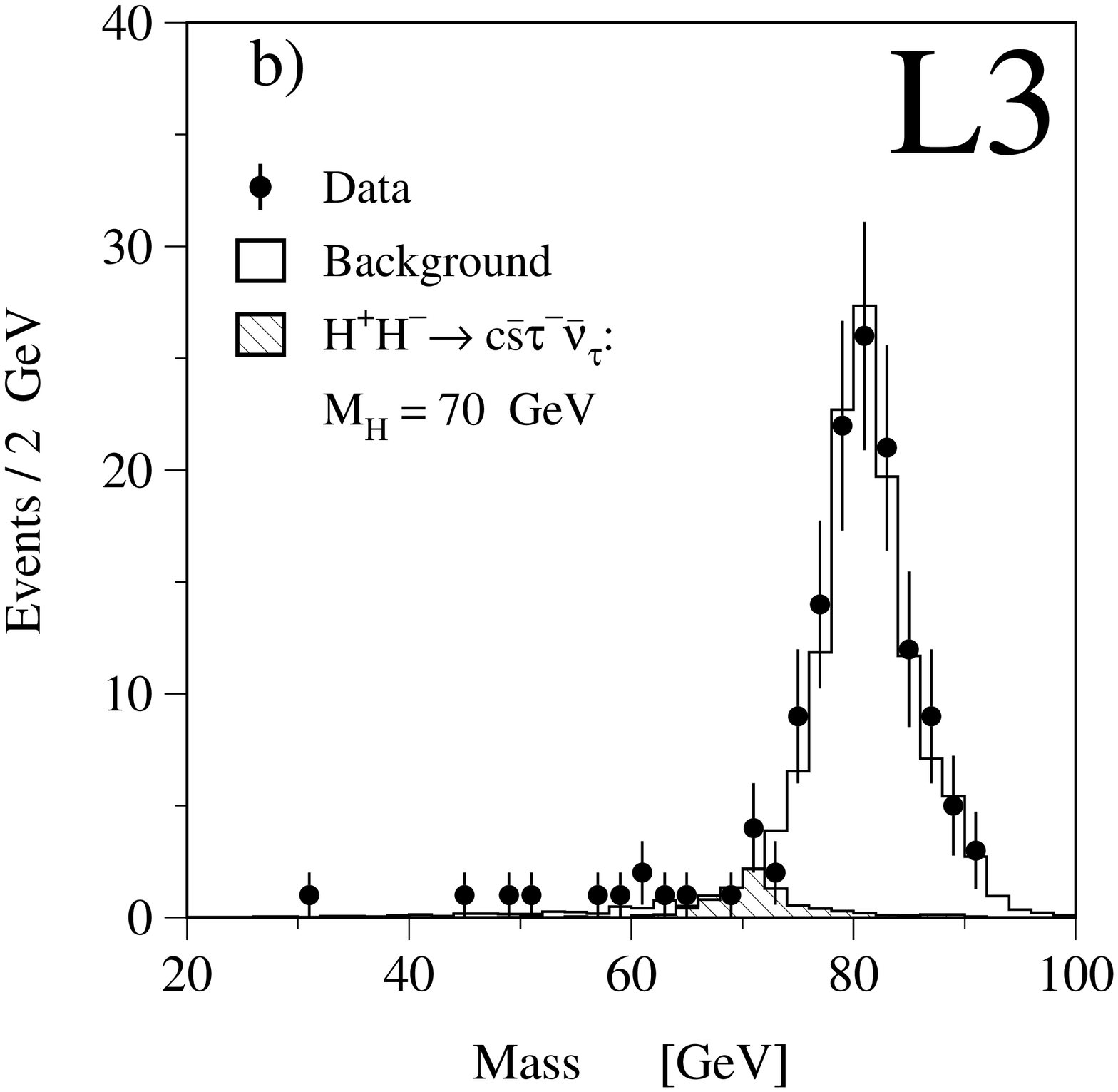,width=0.65\textwidth}} $$
\caption[]{\label{fig:cstn}Distributions for the $ \HHcstn$ channel.  a)
  $\mathrm{E^*_e+|P^*_{miss}}|$  for events with an identified  electron
  in the final  state  after all  other  cuts, the arrow  indicates  the
  position of the cut.  b)  Reconstructed  mass spectrum after all cuts.
  The    expected    distribution    for   a    70~\GeV{}    Higgs    at
  $\mathrm{Br(H^\pm\ra \tau\nu)} = 0.5$ is superimposed, multiplied by a
  factor five in a).}
\end{figure}

High  multiplicity  events  are  selected  by  requiring  more than five
charged tracks, more than 10 calorimetric  clusters and a visible energy
greater  than  $0.35\sqrt{s}$.  The visible mass of the  \mbox{jet--jet}
system must be less than 110 \GeV{} and the energy of the visible  decay
products  of the  tau  lepton  less  than 50  \GeV{}  to  reduce  \qqbar
($\gamma$)   contamination.  This  background  is  further   reduced  by
requiring  the  missing  transverse  momentum  to exceed 25 \GeV{},  the
missing  momentum  parallel to the beam axis to be less than 50\% of the
visible  energy and the  direction  of the  missing  momentum  vector to
satisfy  $|\cos\!\theta_{\mathrm{miss}}| < 0.9$.  The sum of the opening
angles  of the tau  candidate  and the  missing  momentum  vector to the
closest  jet must be larger than  $70^\circ$.  Semileptonic  decays of W
pairs  in  electrons  or  muons  are   suppressed   requiring   the  sum
$\mathrm{E}^*_\l+|\mathrm{P}^*_{\mathrm{miss}}|$  to  be  less  than  60
\GeV{} for an  electron  and less than 50 \GeV{} for a muon in the final
state, where  $\mathrm{E}^*_\l$ and  $\mathrm{P}^*_{\mathrm{miss}}$  are
the energy of the lepton and the missing  momentum  in the rest frame of
the    parent    particle,    respectively.    The    distribution    of
$\mathrm{E}^*_\l+|\mathrm{P}^*_{\mathrm{miss}}|$     is     shown     in
Figure~\ref{fig:cstn}a.    Contamination    from    \qqbar~($\!\gamma$),
two--photon  interactions  and W  pair  events  is  further  reduced  by
requiring $\mid  \cos\!\Theta\mid  \le 0.9$, where $\Theta$ is the polar
angle of the parent particle.

The gain in efficiency with these cuts relative to the analysis at lower
centre--of--mass  energies is 30\% for a Higgs of  80~\GeV{}  mass.  The
selection efficiencies are shown in Table~{\ref{eff}}.  The total number
of events  selected in data is 138, where  129.3  background  events are
expected from Standard Model  processes.  The background is dominated by
the   process   $\Wp\Wm\ra\qqbar^\prime\tau\nu$.  Figure~\ref{fig:cstn}b
shows the average of the masses of the  jet--jet  and the  $\tau$--$\nu$
systems,  calculated  after a kinematic fit imposing energy and momentum
conservation  for  an  assumed  production  of  a  pair  of  equal  mass
particles.

\begin{table}
\begin{center}
\begin{tabular}{|l|ccccc|r|r|} \hline
\multicolumn{1}{|c|}{Channel} & \multicolumn{5}{c|}{Efficiency (\%) for $\mathrm{m_\Hpm}=$} & \multicolumn{1}{c|}{Background} & \multicolumn{1}{c|}{DATA}\\
\multicolumn{1}{|c|}{}        & 60 \GeV  & 65 \GeV & 70 \GeV & 75 \GeV  & 80 \GeV  &               &                     \\ \hline
$\HHtntn$                     &   24     &   27    & 31      &  32      &  34      &  32.5 $\quad$ & $\quad$  30 $\quad$ \\
$\HHcstn$                     &   44     &   42    &  40     &  39      &  34      & 129.3 $\quad$ & $\quad$ 138 $\quad$ \\
$\HHcscs$                     &   39     &   38    &  38     &  34      &  30      & 359.4 $\quad$ & $\quad$ 348 $\quad$ \\ \hline
\end{tabular}
\caption{The  charged Higgs  efficiencies  for various Higgs masses, the
    background  expectations  and the number of data  events.  The total
    error  on the  efficiency  and  on the  background  expectations  is
    estimated to be 5\% and 10\%, respectively.}
\label{eff}
\end{center}
\end{table}

\subsection*{{\boldmath{Search in the $ \HHcscs$ Channel}}}

Events  from the  $\HHcscs$  channel  have a high  multiplicity  and are
balanced in transverse  and  longitudinal  momenta.  A large fraction of
the  centre--of--mass  energy is deposited in the detector, typically by
four hadronic jets.  The  performance of the analysis is unchanged  with
respect to our previous  analysis~\cite{chhiggs_130_180}, hence the same
set of selection cuts is used.

The  selection  efficiencies  are  shown in  Table~\ref{eff}.  The total
number of events selected in data is 348, where 359.4 background  events
are expected from Standard Model  processes.  The main  contribution  to
the   background   comes   from  W   pair   decays   into   four   jets.
Figure~\ref{fig:cscs} shows the average dijet mass after a kinematic fit
imposing  four--momentum  conservation  and equal dijet masses.  The low
mass tail from the $\Wp\Wm$  background is due to  incorrectly  assigned
jet pairs.


\section*{Results}

\begin{figure}[hp]
\centerline{\epsfig{figure=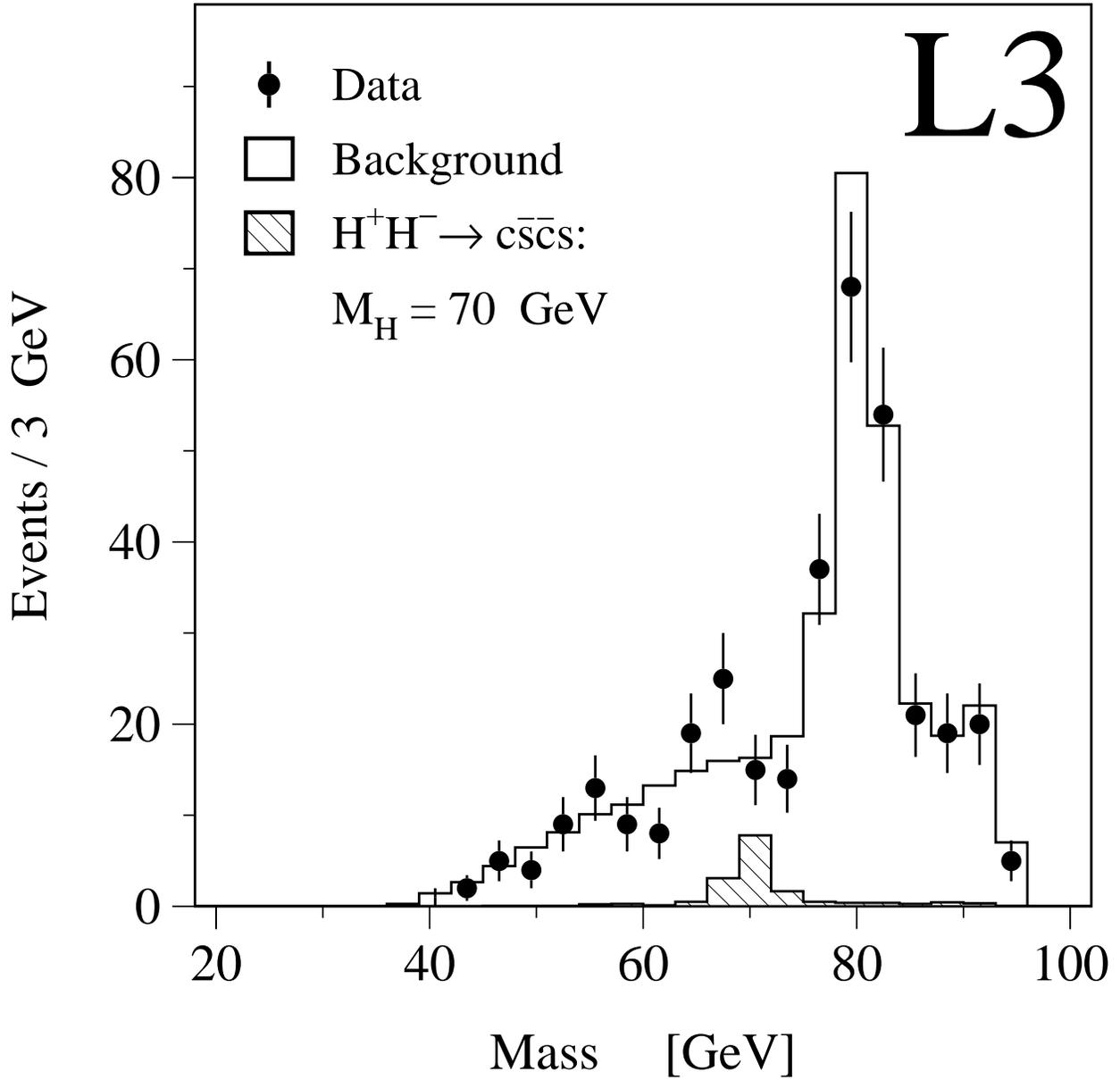,width=1.0\textwidth}}
\caption[]{\label{fig:cscs}  Distribution  of the mass  resulting from a
  kinematic  fit,  with  assumed  production  of a pair  of  equal  mass
  particles,  for data and background  events in the $\HHcscs$  channel.
  The  hatched  histogram  indicates  the  expected  distribution  for a
  70~\GeV{} Higgs at $\mathrm{Br(H^\pm\ra\tau\nu)} = 0$.}
\end{figure}

The number of selected  events in each decay channel is consistent  with
the number of events  expected from  Standard  Model  processes,  and no
significant  deviations in the mass spectra are present.  No  indication
of  pair--produced  charged  Higgs bosons is observed.  Mass limits as a
function of the branching fraction  $\mathrm{Br(H^\pm\rightarrow \tau\nu
)}$ are  derived  at the 95\%  confidence  level~(CL),  using  the  same
technique  described in  Reference~\cite{l3_127}.  For the $\HHcscs$ and
the   $\HHcstn$   channels   the   reconstructed   mass    distributions
(Figures~\ref{fig:cstn}b  and~\ref{fig:cscs})  are  used  in  the  limit
calculation, whereas for the $\HHtntn$ channel the total number of data,
expected background and expected signal events are used.

The systematic  uncertainties on the background and signal are estimated
to be 10\% and  5\%~\cite{sm_189}  respectively,  and come  mainly  from
normalisation  errors due to selection  efficiencies and cross sections.
These errors are included in the confidence level calculation.

\begin{figure}[hp]
\centerline{\epsfig{figure=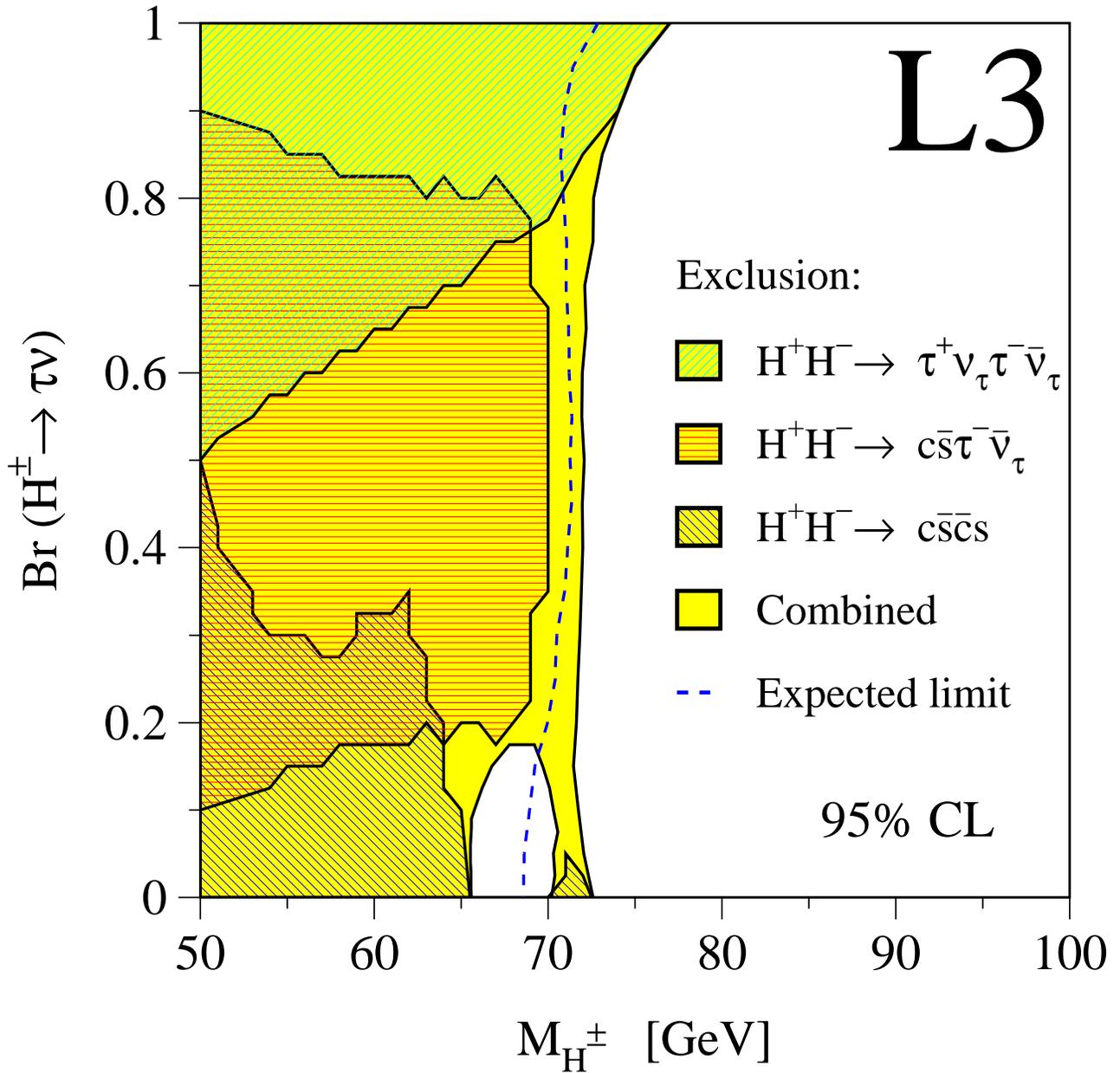,width=1.0\textwidth}}
\caption[]{\label{exclusion}  Excluded  regions  for the  charged  Higgs
  boson   at  95\%  CL  in  the   plane   of  the   branching   fraction
  $\mathrm{Br(H^\pm\rightarrow  \tau\nu)}$ versus mass.  The dashed line
  indicates the median expected exclusion in the absence of a signal.}
\end{figure}

Figure~\ref{exclusion}  shows the excluded mass regions of charged Higgs
bosons  at  95\% CL for the  analyses  of each  final  state  and  their
combination     as     function     of    the     branching     fraction
$\mathrm{Br(H^\pm\ra\tau\nu)}$,  including  the data  from  $\sqrt{s}  =
188.6$~\GeV{}  as well as those from  lower  centre--of--mass  energies.
The  region  around  $\mathrm{m_{H^\pm}}  = 67  \GeV$ at low  values  of
$\mathrm{Br(H^\pm\ra\tau\nu)}$  can only be  excluded at 88\% CL, due to
the slight  excess in data in this mass region  (Figure~\ref{fig:cscs}).
For a branching  fraction of  $\mathrm{Br(H^\pm\ra\tau\nu)}  > 0.2$, the
95\% CL lower limit on the charged Higgs mass is 71.6 GeV.

The limit reveals a significant  improvement on the  sensitivity  of the
data as compared to that from lower centre--of--mass  energies.  A lower
limit    on   the    mass    of   the    charged    Higgs    boson    of
$$\mathrm{m_{H^\pm}}>65.5  \GeV$$ independent of the branching  fraction
is obtained.

\section*{Acknowledgements}

We wish to express our gratitude to the CERN  accelerator  divisions for
the  excellent  performance  of the  LEP  machine.  We  acknowledge  the
efforts  of  engineers  and  technicians  who have  participated  in the
construction and maintenance of the experiment.

\bibliographystyle{l3stylem}

%
%
\newpage
\input namelist180.tex


 \newpage

~

\end{document}

%% file: namelist180.tex
\typeout{   }     
\typeout{Using author list for paper 180 -?}
\typeout{$Modified: Wed Aug 25 10:07:39 1999 by clare $}
\typeout{!!!!  This should only be used with document option a4p!!!!}
\typeout{   }
%
%
%
%
%
%

\newcount\tutecount  \tutecount=0
\def\tutenum#1{\global\advance\tutecount by 1 \xdef#1{\the\tutecount}}
\def\tute#1{$^{#1}$}
\tutenum\aachen            
\tutenum\nikhef            
\tutenum\mich              
\tutenum\lapp              
\tutenum\basel             
\tutenum\lsu               
\tutenum\beijing           
\tutenum\berlin            
\tutenum\bologna           
\tutenum\tata              
\tutenum\ne                
\tutenum\bucharest         
\tutenum\budapest          
\tutenum\mit               
\tutenum\debrecen          
\tutenum\florence          
\tutenum\cern              
\tutenum\wl                
\tutenum\geneva            
\tutenum\hefei             
\tutenum\seft              
\tutenum\lausanne          
\tutenum\lecce             
\tutenum\lyon              
\tutenum\madrid            
\tutenum\milan             
\tutenum\moscow            
\tutenum\naples            
\tutenum\cyprus            
\tutenum\nymegen           
\tutenum\caltech           
\tutenum\perugia           
\tutenum\cmu               
\tutenum\prince            
\tutenum\rome              
\tutenum\peters            
\tutenum\salerno           
\tutenum\ucsd              
\tutenum\santiago          
\tutenum\sofia             
\tutenum\korea             
\tutenum\alabama           
\tutenum\utrecht           
\tutenum\purdue            
\tutenum\psinst            
\tutenum\zeuthen           
\tutenum\eth               
\tutenum\hamburg           
\tutenum\taiwan            
\tutenum\tsinghua          
{
\parskip=0pt
\noindent
{\bf The L3 Collaboration:}
\ifx\selectfont\undefined
 \baselineskip=10.8pt
 \baselineskip\baselinestretch\baselineskip
 \normalbaselineskip\baselineskip
 \ixpt
\else
 \fontsize{9}{10.8pt}\selectfont
\fi
\medskip
\tolerance=10000
\hbadness=5000
\raggedright
\hsize=162truemm\hoffset=0mm
\def\r{\rlap,}
\noindent

M.Acciarri\r\tute\milan\
P.Achard\r\tute\geneva\ 
O.Adriani\r\tute{\florence}\ 
M.Aguilar-Benitez\r\tute\madrid\ 
J.Alcaraz\r\tute\madrid\ 
G.Alemanni\r\tute\lausanne\
J.Allaby\r\tute\cern\
A.Aloisio\r\tute\naples\ 
M.G.Alviggi\r\tute\naples\
G.Ambrosi\r\tute\geneva\
% ...
\newpage
\begin{list}{A}{\itemsep=0pt plus 0pt minus 0pt\parsep=0pt plus 0pt minus 0pt
                \topsep=0pt plus 0pt minus 0pt}
\item[\aachen]
 I. Physikalisches Institut, RWTH, D-52056 Aachen, FRG$^{\S}$\\
 III. Physikalisches Institut, RWTH, D-52056 Aachen, FRG$^{\S}$
\item[\nikhef] National Institute for High Energy Physics, NIKHEF, 
     and University of Amsterdam, NL-1009 DB Amsterdam, The Netherlands
\item[\mich] University of Michigan, Ann Arbor, MI 48109, USA
\item[\lapp] Laboratoire d'Annecy-le-Vieux de Physique des Particules, 
     LAPP,IN2P3-CNRS, BP 110, F-74941 Annecy-le-Vieux CEDEX, France
\item[\basel] Institute of Physics, University of Basel, CH-4056 Basel,
     Switzerland
\item[\lsu] Louisiana State University, Baton Rouge, LA 70803, USA
\item[\beijing] Institute of High Energy Physics, IHEP, 
  100039 Beijing, China$^{\triangle}$ 
\item[\berlin] Humboldt University, D-10099 Berlin, FRG$^{\S}$
\item[\bologna] University of Bologna and INFN-Sezione di Bologna, 
     I-40126 Bologna, Italy
\item[\tata] Tata Institute of Fundamental Research, Bombay 400 005, India
\item[\ne] Northeastern University, Boston, MA 02115, USA
\item[\bucharest] Institute of Atomic Physics and University of Bucharest,
     R-76900 Bucharest, Romania
\item[\budapest] Central Research Institute for Physics of the 
     Hungarian Academy of Sciences, H-1525 Budapest 114, Hungary$^{\ddag}$
\item[\mit] Massachusetts Institute of Technology, Cambridge, MA 02139, USA
\item[\debrecen] Lajos Kossuth University-ATOMKI, H-4010 Debrecen, Hungary$^\P$
\item[\florence] INFN Sezione di Firenze and University of Florence, 
     I-50125 Florence, Italy
\item[\cern] European Laboratory for Particle Physics, CERN, 
     CH-1211 Geneva 23, Switzerland
\item[\wl] World Laboratory, FBLJA  Project, CH-1211 Geneva 23, Switzerland
\item[\geneva] University of Geneva, CH-1211 Geneva 4, Switzerland
\item[\hefei] Chinese University of Science and Technology, USTC,
      Hefei, Anhui 230 029, China$^{\triangle}$
\item[\seft] SEFT, Research Institute for High Energy Physics, P.O. Box 9,
      SF-00014 Helsinki, Finland
\item[\lausanne] University of Lausanne, CH-1015 Lausanne, Switzerland
\item[\lecce] INFN-Sezione di Lecce and Universit\'a Degli Studi di Lecce,
     I-73100 Lecce, Italy
\item[\lyon] Institut de Physique Nucl\'eaire de Lyon, 
     IN2P3-CNRS,Universit\'e Claude Bernard, 
     F-69622 Villeurbanne, France
\item[\madrid] Centro de Investigaciones Energ{\'e}ticas, 
     Medioambientales y Tecnolog{\'\i}cas, CIEMAT, E-28040 Madrid,
     Spain${\flat}$ 
\item[\milan] INFN-Sezione di Milano, I-20133 Milan, Italy
\item[\moscow] Institute of Theoretical and Experimental Physics, ITEP, 
     Moscow, Russia
\item[\naples] INFN-Sezione di Napoli and University of Naples, 
     I-80125 Naples, Italy
\item[\cyprus] Department of Natural Sciences, University of Cyprus,
     Nicosia, Cyprus
\item[\nymegen] University of Nijmegen and NIKHEF, 
     NL-6525 ED Nijmegen, The Netherlands
\item[\caltech] California Institute of Technology, Pasadena, CA 91125, USA
\item[\perugia] INFN-Sezione di Perugia and Universit\'a Degli 
     Studi di Perugia, I-06100 Perugia, Italy   
\item[\cmu] Carnegie Mellon University, Pittsburgh, PA 15213, USA
\item[\prince] Princeton University, Princeton, NJ 08544, USA
\item[\rome] INFN-Sezione di Roma and University of Rome, ``La Sapienza",
     I-00185 Rome, Italy
\item[\peters] Nuclear Physics Institute, St. Petersburg, Russia
\item[\salerno] University and INFN, Salerno, I-84100 Salerno, Italy
\item[\ucsd] University of California, San Diego, CA 92093, USA
\item[\santiago] Dept. de Fisica de Particulas Elementales, Univ. de Santiago,
     E-15706 Santiago de Compostela, Spain
\item[\sofia] Bulgarian Academy of Sciences, Central Lab.~of 
     Mechatronics and Instrumentation, BU-1113 Sofia, Bulgaria
\item[\korea] Center for High Energy Physics, Adv.~Inst.~of Sciences
     and Technology, 305-701 Taejon,~Republic~of~{Korea}
\item[\alabama] University of Alabama, Tuscaloosa, AL 35486, USA
\item[\utrecht] Utrecht University and NIKHEF, NL-3584 CB Utrecht, 
     The Netherlands
\item[\purdue] Purdue University, West Lafayette, IN 47907, USA
\item[\psinst] Paul Scherrer Institut, PSI, CH-5232 Villigen, Switzerland
\item[\zeuthen] DESY, D-15738 Zeuthen, 
     FRG
\item[\eth] Eidgen\"ossische Technische Hochschule, ETH Z\"urich,
     CH-8093 Z\"urich, Switzerland
\item[\hamburg] University of Hamburg, D-22761 Hamburg, FRG
\item[\taiwan] National Central University, Chung-Li, Taiwan, China
\item[\tsinghua] Department of Physics, National Tsing Hua University,
      Taiwan, China
\item[\S]  Supported by the German Bundesministerium 
        f\"ur Bildung, Wissenschaft, Forschung und Technologie
\item[\ddag] Supported by the Hungarian OTKA fund under contract
numbers T019181, F023259 and T024011.
\item[\P] Also supported by the Hungarian OTKA fund under contract
  numbers T22238 and T026178.
\item[$\flat$] Supported also by the Comisi\'on Interministerial de Ciencia y 
        Tecnolog{\'\i}a.
\item[$\sharp$] Also supported by CONICET and Universidad Nacional de La Plata,
        CC 67, 1900 La Plata, Argentina.
\item[$\diamondsuit$] Also supported by Panjab University, Chandigarh-160014, 
        India.
\item[$\triangle$] Supported by the National Natural Science
  Foundation of China.
\item[\dag] Deceased.
\end{list}
}
\vfill



